\documentclass[preprint,aps]{revtex4}
\usepackage{color}

\usepackage{graphicx}

\begin{document}

\title{Observation of a Three-Dimensional Quasi-Long-Range Electronic Supermodulation in YBa$_2$Cu$_3$O$_7$$_-$$_x$/La$_0$$_.$$_7$Ca$_0$$_.$$_3$MnO$_3$ Heterostructures}
\author{Junfeng He$^{1,}$\footnote[1]{These authors contributed equally to this work.}, Padraic Shafer$^{2,*}$, Thomas R. Mion$^{1,*}$, Vu Thanh Tra$^{3}$, Qing He$^{4}$, J. Kong$^{1}$, Y.-D. Chuang$^{2}$, W. L. Yang$^{2}$, M. J. Graf$^{1}$, J.-Y. Lin$^{2,3}$, Y.-H. Chu$^{5,6}$, E. Arenholz$^{2}$, and Rui-Hua He$^{1,\sharp}$}

\affiliation{
	\\$^{1}$Department of Physics, Boston College, Chestnut Hill, MA 02467, USA
	\\$^{2}$Advanced Light Source, Lawrence Berkeley National Laboratory, Berkeley, CA 94720, USA
	\\$^{3}$Institute of Physics, National Chiao Tung University, Hsinchu 30010, Taiwan
	\\$^{4}$Department of Physics, Durham University, Durham DH13LE, UK
	\\$^{5}$Institute of Physics, Academia Sinica, Taipei 11529, Taiwan
	\\$^{6}$Department of Materials Science and Engineering, National Chiao Tung University, Hsinchu 30010, Taiwan
}
%
%



\maketitle

Recent developments in high-temperature superconductivity highlight a generic tendency of the cuprates to develop competing electronic (charge) supermodulations. While coupled to the lattice and showing different characteristics in different materials, these supermodulations themselves are generally conceived to be quasi-two-dimensional, residing mainly in individual CuO$_2$ planes, and poorly correlated along the c-axis. Here we observed with resonant elastic x-ray scattering a distinct type of electronic supermodulation in YBa$_2$Cu$_3$O$_{7-x}$ (YBCO) thin films grown epitaxially on La$_{0.7}$Ca$_{0.3}$MnO$_3$ (LCMO). This supermodulation has a periodicity nearly commensurate with four lattice constants in-plane, eight out-of-plane, with long correlation lengths in three dimensions. It sets in far above the superconducting transition temperature and competes with superconductivity below this temperature for electronic states predominantly in the CuO$_2$ plane. Our finding sheds new light on the nature of charge ordering in cuprates as well as a reported long-range proximity effect between superconductivity and ferromagnetism in YBCO/LCMO heterostructures.

\newpage

A key feature of unconventional superconductors is the presence of multiple ordering tendencies, with some competing (and often coexisting) with superconductivity for electronic states otherwise available for pairing. A case quite often encountered is that these competing orders are affected by, if not entirely driven by, the same underlying interactions that mediate superconductivity \cite{IntertwinedOrder}. Elucidating the exact nature of these interactions would thus promise both useful practical guidance to manipulate superconductivity (via effects on the competing orders) and complementary theoretical insights into the microscopic mechanisms for superconductivity.

Remarkable progress has been made in recent years in identifying charge ordering as a generic competing instability within the cuprate superconductors; nevertheless, a unified understanding of its diverse manifestations in different materials is still lacking. In general, the observed charge orders fall into two main categories. One has an in-plane periodicity close to four lattice constants (period-4) and has been found in La-based cuprates \cite{CO_NdLSCO_Tranquada, CO_EuLSCO_Fink, CO_LBCO_Abbamonte}, Ca$_{2-x}$Na$_x$CuO$_2$Cl$_2$ \cite{CO_CCOC}, Bi$_2$Sr$_2$CaCu$_2$O$_{8+\delta}$ (Bi2212)\cite{CO_Bi2212_Davis, CO_Bi2212_Kapitulnik, CO_Bi2212_Yazdani}, YBCO \cite{CO_YBCO_NMR1}, Bi$_2$Sr$_{2-x}$La$_x$CuO$_{6+\delta}$ (Bi2201)\cite{CO_Bi2201_Damascelli} and Nd$_{2-x}$Ce$_x$CuO$_4$ \cite{CO_NCCO}. Another category has a periodicity generally different from four lattice constants that varies in Bi2201 \cite{CO_Bi2201_Hudson}, YBCO \cite{CO_YBCO_Ghiringhelli, CO_YBCO_Chang, CO_YBCO_Achkar, CO_YBCO_Blanco-Canosa, CO_YBCO_Hucker}, Bi2212 \cite{CO_Bi2212_Yazdani} and HgBa$_2$CuO$_{4+\delta}$ \cite{CO_Hg1201}. For bulk YBCO, charge order with periodicity close to three lattice constants (period-3) was observed in resonant elastic x-ray scattering (REXS) and hard x-ray diffraction experiments in zero or relatively low magnetic field. This ordering has an onset at $\sim$ 110-160 K upon cooling and maintains short-range order at even lower temperatures \cite{CO_YBCO_Ghiringhelli, CO_YBCO_Chang, CO_YBCO_Achkar, CO_YBCO_Blanco-Canosa, CO_YBCO_Hucker}. A period-4 charge order was found in nuclear magnetic resonance (NMR) experiments at high fields which has long-range order and develops below $\sim$ 45-60 K in addition to the period-3 order \cite{CO_YBCO_NMR2}. It remains uncertain whether the two categories of order reflect the same charge ordering mechanism in the cuprates that is prone to the effects of various internal (crystalline environment, quenched disorder) and external factors (magnetic field), or whether they should be attributed to distinct tendencies of different origins, each competing with superconductivity \cite{DisorderEffectPNAS}.

In this paper we report the observation of a distinct electronic order in YBCO thin films grown epitaxially on LCMO layers. This order shares similarities with both categories of charge order found in bulk YBCO, yet shows important distinctions. Like the bulk orders, the novel electronic order propagates along the Cu-O bond directions, predominantly involves electronic states in the CuO$_2$ plane, and competes with superconductivity. It has a similar periodicity as the bulk period-4 charge order and also shares a long correlation length in the CuO$_2$ plane, but has a much higher onset temperature, $\sim$ 220 K. These characteristics are different from those of the period-3 bulk charge orders found under similar experimental conditions using the same technique, REXS. Crucially, this electronic order exhibits a remarkable correlation along the crystalline c-axis, which is unprecedented among all the charge orders found so far in the cuprates. We found that the onset of the electronic order does not coincide with that of bulk ferromagnetism in the LCMO layer. Our observations add a significant piece to the puzzle of high-temperature superconductivity, providing new insights into the nature of charge ordering in cuprates. Additionally this three-dimensional (3D) long-range competing order in YBCO/LCMO heterostructures provides a new angle for interpreting the long-range proximity effect between superconductivity and ferromagnetism reported in this system.
\\

\begin{figure*}[tbp]
\begin{center}
\includegraphics[width=0.73\columnwidth,angle=0]{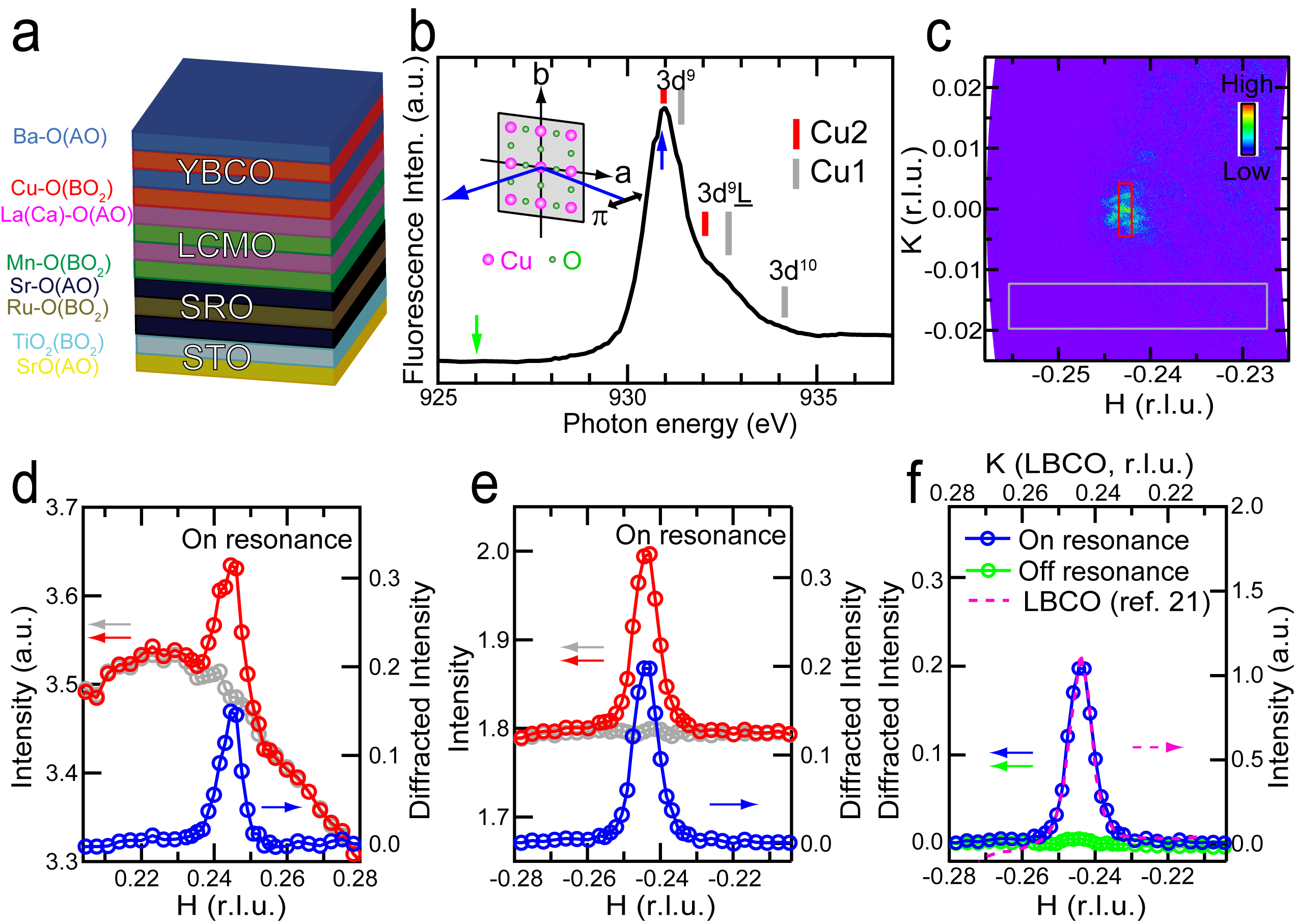}
\end{center}
\vspace*{-0.7cm}
\caption{\textbf{In-plane component of the electronic supermodulation.} \textbf{a}, Schematic of the heterostructure in Sample \#1. The individual atomic layers (AO and BO$_2$) of the ABO$_3$-type perovskite are indicated by different colors. \textbf{b}, X-ray absorption spectrum measured near the Cu $L_3$ absorption edge in total fluorescence yield. Main configuration-dependent contributions to the spectrum are marked by red and grey bars as reproduced from Ref. \cite{CO_YBCO_Ghiringhelli}. Red and grey bars mark the energy positions of states derived from the Cu sites in the CuO$_2$ plane (Cu2) and those in the Cu-O chain layer (Cu1), respectively. Inset: Schematic of the scattering geometry with linear ($\pi$) incident polarization in the horizontal scattering plane. \textbf{c}, In-plane reciprocal space projection derived from a single CCD image measured at 930.85 eV (indicated by blue arrow in \textbf{b}) showing a diffraction peak near $(-0.245,0,1.38)$. Note that the raw image contains the $L$ range from $\sim$1.31 to $\sim$1.45 and only its projection onto the $(H, K, 0)$ plane is shown here. \textbf{d} \& \textbf{e}, Scattered intensity line cuts along $H$ measured at 50 K and 930.85 eV at fixed $L=1.38$ and $K=0$. Each point on the red (grey) curve indicates average intensity in the region of the corresponding CCD image marked by the red (grey) rectangle in \textbf{c}. Blue curve is the difference between red and grey curves showing the diffracted intensity. The intensity in \textbf{e} and diffracted intensity in \textbf{d}-\textbf{f} all share the same scale with the intensity in \textbf{d}. \textbf{f}, Diffracted intensity measured at 926 eV (green curve, indicated by the green arrow in \textbf{b}) and 930.85 eV (blue curve, reproduced from \textbf{e}) at fixed $L=1.38$ and $K=0$, compared alongside the $K$ scan of La$_{1.875}$Ba$_{0.125}$CuO$_4$ (reproduced from Ref. \cite{CO_YBCOvsLBCO}; note that equivalent modulations exist along both the $H$ and $K$ directions in the low-temperature tetragonal phase of La$_{1.875}$Ba$_{0.125}$CuO$_4$). All results were obtained from Sample \#1.}
\label{Fig. 1}
\end{figure*}

\noindent{\bf\large Results}\\
\noindent{\bf In-plane component of the electronic supermodulation.}
REXS combines the periodic spatial information of x-ray diffraction with the sensitivity to local electronic bonding environment found in soft x-ray absorption fine structure. By using photons with energy near the Cu $L_3$ absorption edge ($2p_{3/2}\rightarrow 3d$ transition) [Fig. \ref{Fig. 1}b], REXS can directly detect the ordering of valence electronic states in cuprates that involves modulations in their charge density and/or band energy. Earlier experiments have shown that REXS is particularly sensitive to charge order \cite{CO_YBCO_Ghiringhelli, CO_YBCO_Achkar, CO_YBCO_Blanco-Canosa}.
\\
The YBCO thin films were grown epitaxially on LCMO layers (Supplementary Fig. 1). The intensity of x-rays scattered from YBCO/LCMO films (Fig. \ref{Fig. 1}a) were measured as a function of momentum transfer along primary crystallographic directions using photons both on resonance (930.85 eV) and off resonance (926 eV). Diffraction spots (Fig. \ref{Fig. 1}c) were observed with resonant photons at symmetric points in reciprocal space, $(\pm \emph{H, 0, L})$, where indices $H$ and $L$ indicate scaling of the reciprocal lattice units (r.l.u.), $2\pi/a$ and $2\pi/c$, respectively. Line cuts through the diffraction peaks along the in-plane Cu-O bond direction ($a$-axis) are shown for $(0.245, 0, 1.38)$ [Fig. \ref{Fig. 1}d] and $(-0.245, 0, 1.38)$ [Fig. \ref{Fig. 1}e]. The on-resonance diffraction peaks are superimposed upon a diffuse background of comparable intensity. The diffuse background contains components of x-ray fluorescence from the sample as well as scattered x-rays that do not contribute to the diffraction signal. As shown in Figs. \ref{Fig. 1}d \& e, the on-resonance diffuse background--measured at detector angles slightly away from the diffraction condition--closely matches the shape of the scattered intensity line cuts everywhere except in close proximity to $H= \pm 0.245$. By subtracting the diffuse background, the true on-resonance diffraction peaks are revealed (also shown in Figs. \ref{Fig. 1}d \& e).

These diffraction peaks are located at $|H|= 0.245 \pm 0.005$ (the error bars come from the uncertainties in the determination of the peak positions), very close to the commensurate wavevector $1/4$, indicative of a supermodulation having a period close to $4a$ ($a=3.855$ $\AA$ \cite{CO_YBCO_Ghiringhelli}) along the Cu-O bond direction. This periodicity is different from those of the period-3 charge orders found in the bulk YBCO by REXS with wavevector $\sim0.31-0.34$ \cite{CO_YBCO_Ghiringhelli, CO_YBCO_Chang, CO_YBCO_Achkar, CO_YBCO_Blanco-Canosa, CO_YBCO_Hucker}, but consistent with the one inferred from a NMR study \cite{CO_YBCO_NMR1}. The observed diffraction peak has a full width half maximum (FWHM) $\delta H=0.0075$, corresponding to a correlation length $\xi_{ab}\doteq a(\pi\delta H)^{-1} \sim 164$ $\AA$ ($\sim$ 42 unit cells, u.c.) for the supermodulation, which is two times larger than the longest correlation length reported on the period-3 charge orders in bulk YBCO \cite{CO_YBCO_Blanco-Canosa, CO_YBCO_Hucker}. We note that both the periodicity and correlation length of the supermodulation in YBCO/LCMO are quantitatively similar to those of the long-range charge stripe order in La$_{1.875}$Ba$_{0.125}$CuO$_4$ \cite{CO_YBCOvsLBCO} (Fig. \ref{Fig. 1}f). Moreover, we observed that the supermodulation diffraction peaks diminish rapidly away from resonance (Fig. \ref{Fig. 1}f). Such a strong resonant effect indicates a sizable electronic character of the supermodulation that necessarily involves Cu $3d$ valence states.
\\

\begin{figure*}[tbp]
\begin{center}
\includegraphics[width=0.7\columnwidth,angle=0]{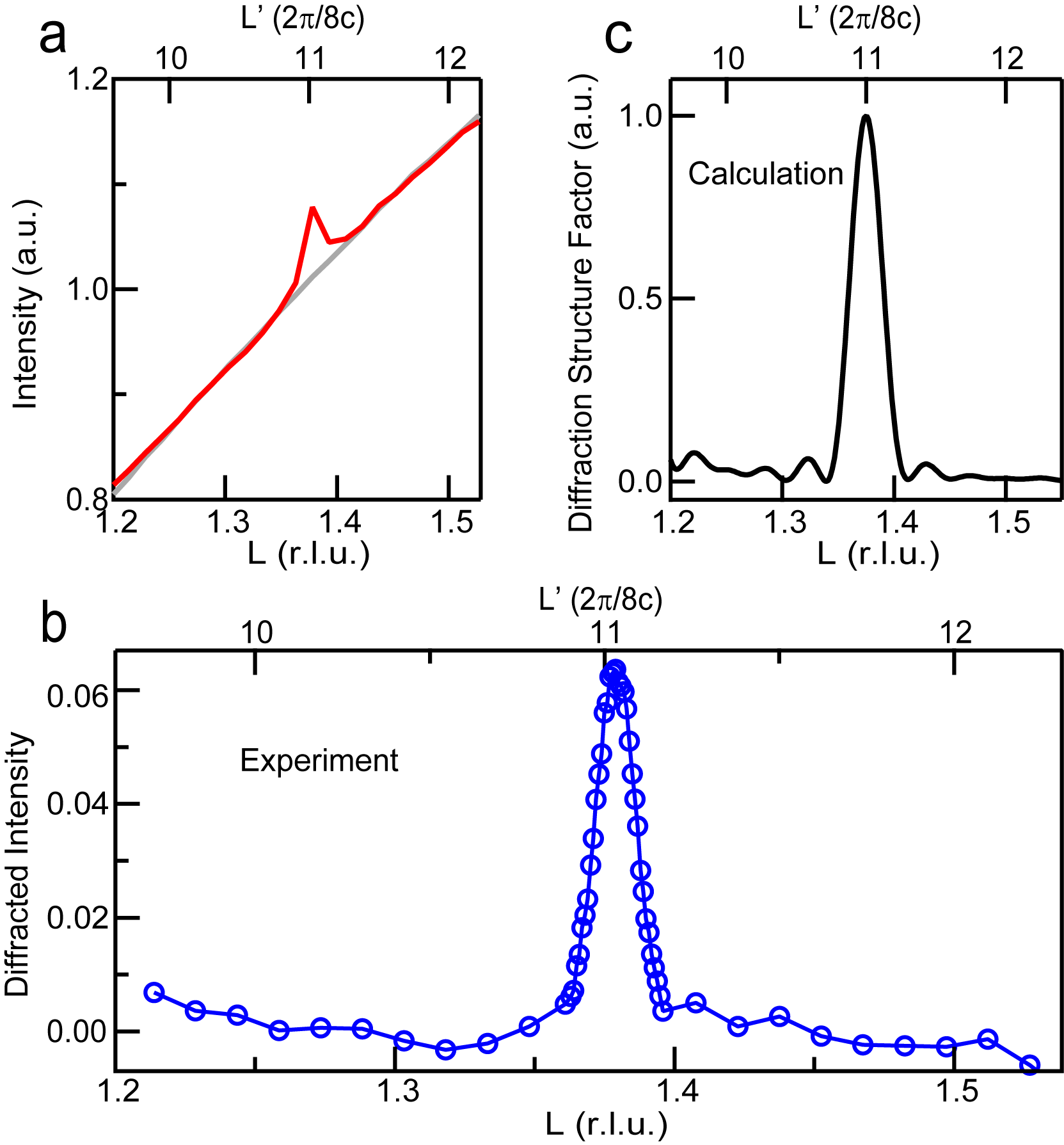}
\end{center}
\caption{\textbf{Out-of-plane component of the electronic supermodulation.} \textbf{a}, Scattered intensity line cuts along $L$ measured on Sample \#1 at 50 K and 930.6 eV at fixed $H=-0.245$ and $K=0$. Average intensity in the peak region and background region of the CCD image is shown in red and grey curves, respectively. \textbf{b}, Diffracted intensity for the same line cut which is the difference between red and grey curves in \textbf{a}. The diffracted intensity in \textbf{b} shares the same scale with the intensity in \textbf{a}. \textbf{c}, Calculated diffraction structure factor for one representative type of charge density modulation along the c-axis.}
\label{Fig. 2}
\end{figure*}

\noindent{\bf Out-of-plane component of the electronic supermodulation.}
Fig. \ref{Fig. 2}a shows the scattered intensity and nearby background intensity as a function of $L$ at $(-0.245, 0, L)$. Their difference is plotted in Fig. \ref{Fig. 2}b to reveal the true diffraction peak, similar to the line cuts along $H$ shown in Fig. \ref{Fig. 1}. The diffracted intensity exhibits a strong $L$-dependence, dominated by a prominent peak with FWHM $\delta L=0.017$. This suggests a remarkable correlation of the supermodulation along the c-axis with a correlation length $\xi_c\doteq c(\pi\delta L)^{-1} \sim 219$ $\AA$ ($\sim$ 19 u.c.; $c=11.64$ $\AA$). This result strongly contrasts with the weak $L$-dependent modulations of the diffracted intensity and the short c-axis correlation length ($\sim$ 1 u.c.) of the charge orders found in various bulk cuprates by REXS \cite{CO_LBCO_Abbamonte, CO_YBCO_Ghiringhelli, CO_Bi2201_Damascelli}. It points to a 3D quasi-long-range order of the electronic supermodulation in YBCO/LCMO.

The diffraction peak is located at $L=1.38 \pm 0.02$, very close to the commensurate wavevector $L=11/8$, hinting at an approximate $8c$ period of the supermodulation along the c-axis. Within the experimentally accessible $L$ range, no diffraction peak is found near commensurate wavevectors $L=10/8$ and $L=12/8$. As shown in Supplementary Fig. 2, we have calculated diffraction structure factors for four representative types of period-8 charge density modulation along the c-axis each with a different phase relation within the CuO$_2$ bilayer as well as between neighboring bilayers. A common feature of these calculations is that diffraction peak is present at odd harmonics of $L=1/8$ but not at even harmonics, although the relative peak intensity varies with the form of modulation. Such a robust qualitative aspect of the calculations is consistent with our experimental observation for the out-of-plane component of the electronic supermodulation (cf. Fig. \ref{Fig. 2}b \& c).
\\

\begin{figure*}[tbp]
\begin{center}	
\includegraphics[width=1.0\columnwidth,angle=0]{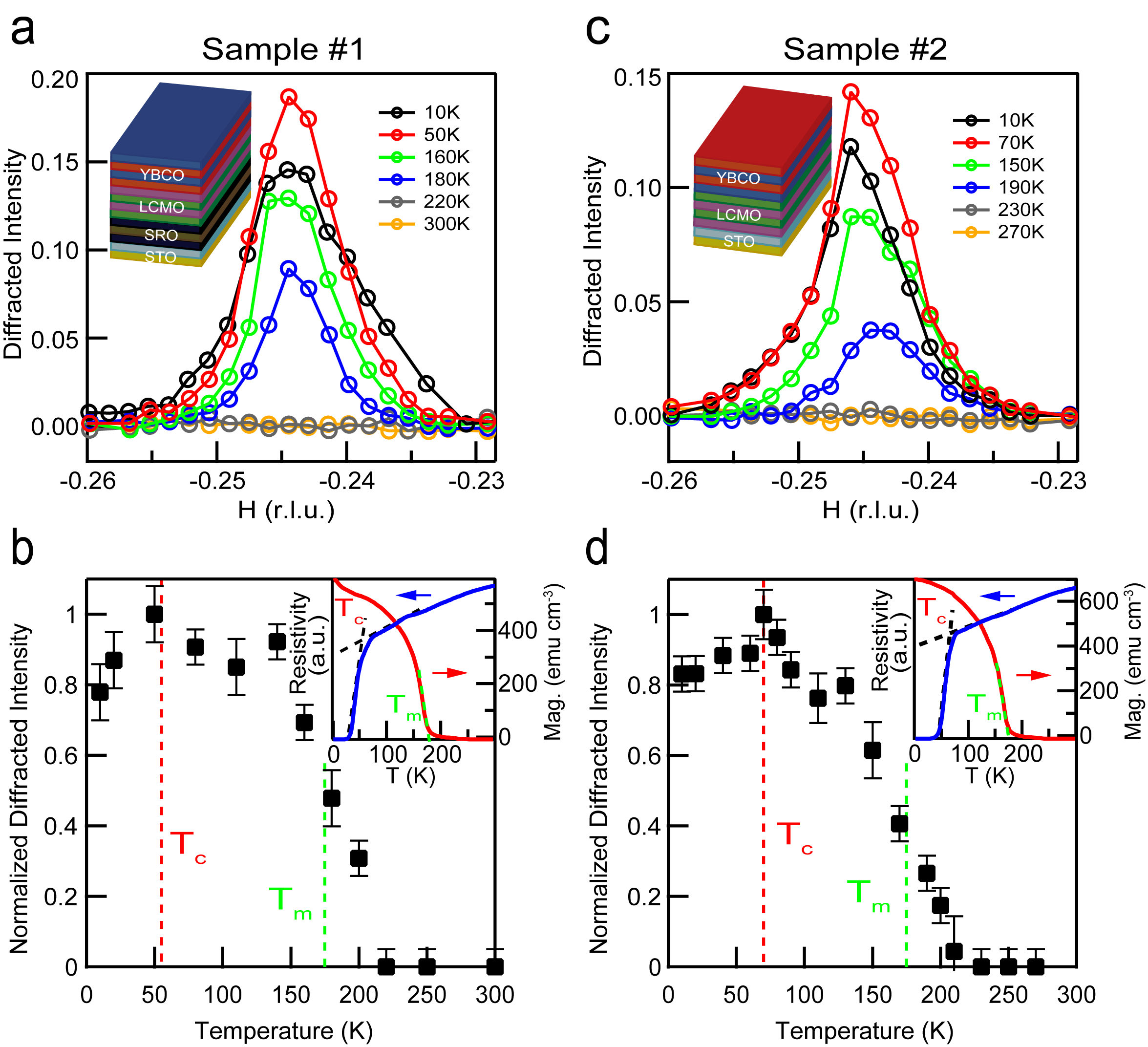}
\end{center}
\caption{\textbf{Temperature evolution of the electronic supermodulation.} \textbf{a}, Diffracted intensity of a line cut along $H$ measured at resonance (930.85 eV) on Sample \#1 at selected temperatures at fixed $L=1.38$ and $K=0$. \textbf{b}, Temperature dependence of the peak intensity (normalized to the maximum value) in \textbf{a}. Red (green) dashed line marks $T_c$ ($T_m$) determined from the resistivity (magnetization) measurement shown in the insets. \textbf{c} \& \textbf{d}, Results of Sample \#2. The diffracted intensity in \textbf{a} \& \textbf{c} share the same scale as before. Insets of \textbf{a} \& \textbf{c} are schematics of the heterostructure for Samples \#1 \& \#2, respectively. Error bars reflect the uncertainty due to mean-square deviation in determining the diffraction peak intensity based on curve fitting.}
\label{Fig. 3}
\end{figure*}

\noindent{\bf Temperature evolution of the electronic supermodulation.}
We have studied the diffraction peak centered at $(-0.245,0,1.38)$ as a function of temperature on the same YBCO/LCMO sample, with results summarized in Fig. \ref{Fig. 3}a \& b. No diffraction peak is found at room temperature. It emerges upon lowering the temperature, showing a clear onset behavior that appears more abruptly than those observed on the period-3 charge orders in bulk YBCO \cite{CO_YBCO_Ghiringhelli, CO_YBCO_Chang, CO_YBCO_Achkar, CO_YBCO_Blanco-Canosa, CO_YBCO_Hucker}. This onset occurs at $T_{co}\sim$ 220 K, substantially higher than the period-3 charge ordering onset $\sim$ 150 K. $T_{co}$ is also significantly higher than the Curie temperature of the LCMO layer $T_m\sim$ 175 K, which was determined by a magnetization measurement on the YBCO/LCMO sample (the red curve in the inset of Fig. \ref{Fig. 3}b).

The diffracted intensity grows at first rapidly upon further lowering the temperature and then more gradually $\sim$ 140 K (Fig. \ref{Fig. 3}b). The intensity reaches a maximum at $T_{max}\sim$ 50 K, below which intensity decreases. During the entire temperature evolution the diffraction peak stays at roughly the same position (Fig. \ref{Fig. 3}a) with only modest changes to the peak width (Supplementary Fig. 3). We note that the superconducting transition temperature of the YBCO layer $T_c$ was determined from a four-probe resistivity measurement on the YBCO/LCMO sample to be 55 $\pm$ 10 K (the blue curve in the inset of Fig. \ref{Fig. 3}b), which is nearly indistinguishable from $T_{max}$ on this sample (Sample \#1).

We have performed similar measurements on a second YBCO/LCMO sample (Sample \#2) that was grown with a different interface termination than Sample \#1 (see Methods and Discussion). Sample \#2 shows a slightly higher $T_c=72 \pm 7$ K than Sample \#1 but a similar $T_m$ (the inset of Fig. \ref{Fig. 3}d). A similar supermodulation was found in this sample, with an overall similar temperature evolution of the diffraction peak (cf. Fig. \ref{Fig. 3}a \& b vs. c \& d). A notable difference is $T_{max}$, which is now $\sim$ 75 K. We emphasize that $T_{max} \approx T_c$ in each sample.

The observed suppression of the diffraction peak below the superconducting transition in YBCO/LCMO is reminiscent of those observed in bulk YBCO \cite{CO_YBCO_Ghiringhelli, CO_YBCO_Chang, CO_YBCO_Achkar, CO_YBCO_Blanco-Canosa, CO_YBCO_Hucker}. Similar temperature-dependent behavior has generally been taken as evidence for a microscopic coexistence and dynamic competition between two distinct orders \cite{Tcsuppresion_pnictide, Tcsuppresion_Bi2212}. Such coexistence can occur in a homogenous state. Nevertheless, a rather typical situation is that the two orders are phase separated at the nanoscale with a volume fraction that changes with temperature. Regardless of which type of spatial coexistence is present, the observed temperature dependence suggests that a considerable portion of electronic states that participated in the electronic supermodulation above $T_c$ are instead consumed by superconductivity below $T_c$. Because the states involved in superconductivity lie primarily within the CuO$_2$-plane, the electronic supermodulation in YBCO/LCMO is accordingly expected to have a sizable contribution from the states in the CuO$_2$ plane.
\\

\noindent{\bf Primary CuO$_2$ plane origin of the electronic supermodulation.}
A period-4 supermodulation in the ab-plane could in principle be caused by the ordering of oxygen into arrays of full and empty chains in the Cu-O chain layer. While such period-4 (aka ortho-IV) oxygen order has, to our knowledge, not been reported in YBCO thin films, it has been observed in bulk YBCO at and above the optimal doping level \cite{IslamPRL, StrempferPRL}. Like other oxygen orderings the ortho-IV order persists well above room temperature (up to 500 K) and has a weak temperature dependence, with no anomaly observed at $T_c$. It remains short-ranged in the ab-plane and poorly correlated along the c-axis, with correlation lengths $\sim$3-6 u.c. and $<$ 1 u.c., respectively. All of these characteristics strongly contrast with our observations in YBCO/LCMO.

\begin{figure*}[tbp]
\begin{center}
\includegraphics[width=1.0\columnwidth,angle=0]{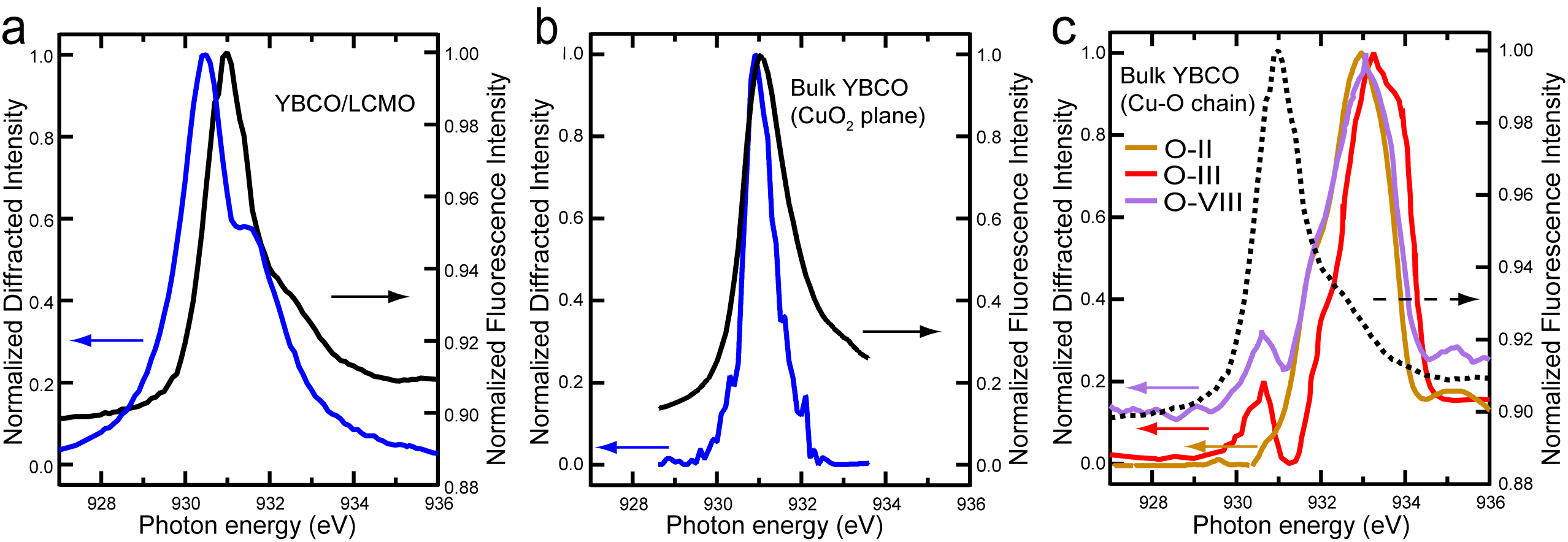}
\end{center}
\caption{\textbf{Comparison on the resonance profile for various orders.} \textbf{a}, Photon energy dependence of the diffracted intensity (normalized to the maximum) at $(-0.245,0,1.38)$ associated with the electronic supermodulation (resonance profile; blue curve), in comparison with that of the fluorescence (x-ray absorption spectrum; black), measured on Sample \#1 at 50 K. \textbf{b}, Resonance profile of the period-3 charge order and x-ray absorption spectrum in a bulk YBCO sample, reproduced from Ref. \cite{CO_YBCO_Achkar}. \textbf{c}, Resonance profiles of various oxygen orderings, ortho-II, ortho-III and ortho-VIII, in different bulk YBCO samples, reproduced from Refs. \cite{CO_YBCO_Achkar, Chain_YBCO_Hawthorn}. Dashed curve is the x-ray absorption spectrum of Sample \#1 reproduced as a guide to the eye. Slight energy offsets in the energy axis between different experiments have been corrected by setting the corresponding $L_3$ absorption peak to the same energy. Note that conditions may vary in different experiments.}
\label{Fig. 4}
\end{figure*}

Our REXS experiment provides further insights into the chain (Cu1) vs. plane (Cu2) origin of the observed electronic supermodulation; here we maintain the copper site (Cu1, Cu2) terminology used in Ref. \cite{CO_YBCO_Ghiringhelli}. By analyzing the photon energy dependence of the diffracted intensity, we directly probe the site, configuration and orbital nature of the states that participate in the associated supermodulation. This resonance profile of the diffraction peak is generally distinct from the x-ray absorption spectrum that is obtained, e.g., from the fluorescence near the Cu $L_3$ edge that involves all unoccupied Cu $3d$ orbitals, regardless of whether they are ordered. As shown in Fig. \ref{Fig. 4}a, the resonance profile measured at $(-0.245,0,1.38)$ in our YBCO/LCMO sample shows a dominant peak close to the $L_3$ absorption maximum at 931 eV. This situation qualitatively resembles the case of the period-3 charge order in bulk YBCO, an example of which is reproduced from the literature in Fig. \ref{Fig. 4}b \cite{CO_YBCO_Achkar}. This peak has been ascribed to a $3d^9$ ground-state configuration at the Cu (Cu2) site in the CuO$_2$ plane (as marked in Fig. \ref{Fig. 1}b) \cite{CO_YBCO_Ghiringhelli, CO_YBCO_Achkar}. In our study, the peak is observed $\sim$ 0.55 eV below the absorption maximum. This finite energy difference between the Cu2 $3d^9$ peak of the resonance profile and the absorption maximum is not uncommon for cuprates, and its size has been found to vary considerably among different experiments \cite{CO_EuLSCO_Fink, CO_LBCO_Abbamonte, CO_YBCO_Achkar, CO_LSCO_Wu} even in the same class of materials. We note that the possible presence of ``dead" layers on the sample surface could account for this energy difference. These surface layers would not contribute to the scattered intensity but would absorb and effectively suppress the peak intensity around the absorption maximum, thereby shifting the apparent diffraction peak toward lower energy. On the other hand, if this energy difference is intrinsic, it may provide a useful constraint on the type of spatial modulation that underlies the electronic supermodulation, e.g., whether it is dominated by direct modulation of the local charge density or else, such as the local Cu $2p\rightarrow 3d$ transition energy \cite{CO_YBCO_Achkar}. In this context, future studies might investigate to what extent the larger energy difference found in YBCO/LCMO versus in the bulk (cf. Fig. \ref{Fig. 4}a vs. b) reflects the characteristics of these distinct electronic structure modulations.
		
We next consider that a shoulder is clearly present in our resonance profile at $\sim$ 1 eV above the main peak (Fig. \ref{Fig. 4}a). Based on this energy separation, we tentatively assign the shoulder feature to the $3d^9\underline{L}$ configuration at the Cu2 site (as marked in Fig. \ref{Fig. 1}b) \cite{CO_YBCO_Ghiringhelli}. Nevertheless, a theoretical calculation that takes into account all possible states in proximity, such as the $3d^9$ configuration at the Cu (Cu1) site in the Cu-O chain layer, might be needed for a concrete understanding of this secondary feature. Most notably the well-defined three-dimensionality of the electronic supermodulation probably involves cooperation with the entire lattice, which would necessitate that additional sites outside of the CuO$_2$ planes contribute to the resonance profile. Its detailed line shape thus holds a key to further elucidate the electronic supermodulation and its distinction from the bulk charge order.

In contrast, our measured resonance profile looks qualitatively different from those for various oxygen orders in the Cu-O chain layer of bulk YBCO, which we reproduce in Fig. \ref{Fig. 4}c \cite{CO_YBCO_Achkar, Chain_YBCO_Hawthorn}. Although the spectral line shape depends on the actual configuration of the oxygen order and the doping level, a common characteristic of these resonance profiles is that the dominant peak is located at least 2 eV above the absorption edge. The relevant states in this energy range are mostly associated with a $d^9\underline{L}$ configuration of the Cu1 site \cite{CO_YBCO_Ghiringhelli} which is directly impacted by the oxygen ordering. Because these features are so distinct from the characteristics we observed, it appears unlikely that the resonance profile of the electronic supermodulation can be consistent with an ordering of oxygen in the chain layer.

Consistent with the temperature dependence results, all of these details gleaned from the resonance profile support a scenario in which the CuO$_2$ planes are the primary locations of the electronic supermodulation in YBCO/LCMO.
\\

\noindent{\bf\large Discussion}\\
What is the nature of the observed electronic supermodulation? Although some unconventional orders have been proposed to exist in cuprates, only 2D charge and spin orders have been found experimentally, with the in-plane charge ordering wavevectors located around $(0,0)$ and the spin ordering wavevectors around $(1/2,1/2)$. In this context, our electronic supermodulation with an in-plane wavevector $(0.245,0)$ is likely a charge order. Our 30-nm YBCO films contain structurally twinned domains and have a pseudo-tetragonal structure, similar to the case in Nd$_{1.2}$Ba$_{1.8}$Cu$_3$O$_7$ (NBCO) epitaxial films \cite{Pseudotetragonal}. This has made it difficult to distinguish between uniaxial and biaxial supermodulations with the typical spot size (100$\times$100 $\mu$m$^2$) of our x-ray beam much larger than the domain size. Because of the relatively small photon momentum at the Cu $L_3$ resonance, the measured intensity line cuts along $H$ and $L$ cover only a limited range in the reciprocal space [e.g., $(-0.245, 0, 13/8)$ is inaccessible] and do not provide sufficient information to reconstruct the full arrangement of charge disproportionation (spatial configuration) of the supermodulation. Future scattering experiments promise further insights into this order, in terms of its charge vs. spin nature by studying the polarization dependence of the diffracted intensity \cite{CO_YBCO_Ghiringhelli}; in terms of the in-plane anisotropy (order along a- vs. b-axis) via spatially-resolved scattering experiments; and in terms of the spatial configuration based on diffraction structure factors determined from measurements over a wider region of the reciprocal space using hard x-rays \cite{CO_IXS_LeTacon}. Despite these open questions, the 3D electronic supermodulation we directly observed is the first of its kind ever reported in cuprates.

To the extent that our observation is indeed charge order, its in-plane periodicity suggests a close link with the period-4 charge order found in bulk YBCO by NMR \cite{CO_YBCO_NMR1}. As a local probe, NMR does not readily distinguish whether an order is 2D or 3D in layered materials, although its finding on YBCO has been interpreted by assuming a 2D order typical for cuprates. In light of the ubiquity of period-4 charge orders in various cuprate families, it is not implausible that both orders reflect the same inherent tendency toward period-4 charge ordering. In such a scenario, possible variations in the dimensionality, correlation length, and onset temperature of the orders may reflect the susceptibility of such tendency to various internal (strain, quenched disorder) and external factors (crystalline environment, magnetic field) as recently discussed \cite{DisorderEffectPNAS}.

However, because two different types of (period-3 and period-4) electronic supermodulation have been observed in nominally the same material (YBCO) with the same technique (REXS) under similar experimental conditions, a new question is posed: What is the possible cause for such a difference? To pinpoint this cause would be the first step to understand whether or not both types of supermodulation should be attributed to distinct tendencies of different origins \cite{DisorderEffectPNAS}.

We first consider the carrier concentration. Determining and controlling the oxygen content in oxide thin film growth is still a general experimental challenge. While our deposition and annealing recipe is optimized for the growth of an optimally-doped YBCO thin film directly on the SrTiO$_3$ (001) substrate (see Methods), the overall oxygen content of the YBCO film that is instead grown on the intermediate LCMO layer could likely shift the effective doping away from the optimal level, given the different crystalline environment during the growth. Moreover, charge transfer at the YBCO/LCMO interface tends to decrease the overall hole concentration of the YBCO layer \cite{Chakhalian_Science}. It was recently found \cite{Tra} that the charge transfer is stronger at the La$_{0.7}$Ca$_{0.3}$O-terminated interface (associated with our Sample \#1 of $T_c=55$ K) than at the MnO$_2$-terminated interface (associated with our Sample \#2 of $T_c=72$ K). It is thus likely that both the YBCO thin films in Samples \#1 \& \#2 with period-4 supermodulations present are located in the underdoped regime, where period-3 charge orders were observed in the bulk. Therefore, carrier concentration may not be essential to account for the aforementioned difference.

Given the magnetic properties of LCMO, we must certainly consider their effect on charge order. We found that $T_{co}$ is 45 K higher than $T_m$, which marks the onset of macroscopic ferromagnetism in the LCMO layer. Ferromagnetism has been found to set in at the YBCO/LCMO interface slightly below $T_m$ \cite{Chakhalian_NP}, in contrast to the temperature of $T_{co}$ that we observe. Therefore, the long-range ferromagnetism of LCMO does not appear to be directly relevant to the formation of the electronic supermodulation. While short-range ferromagnetism may possibly persist above $T_m$, it is unlikely to play a major role in light of the abrupt onset at $T_{co}$ and the lack of anomaly at $T_m$ observed in the temperature evolution of the electronic supermodulation. In bulk YBCO, we note that the charge ordering onset is independent of external magnetic field. But magnetic field suppresses superconductivity and can consequently enhance the charge order near and below $T_c$ \cite{CO_YBCO_Chang}. In the case of YBCO/LCMO, ferromagnetism at the interface or induced in the YBCO layer may exert a similar indirect effect on the electronic supermodulation via the exchange field. Whether the two can couple in a different way remains to be explored.

Therefore we turn our attention to the effects of strain. Strain can affect the energetics of competing ordering tendencies or the forms in which a given tendency manifest by tipping the balance one way or the other. Compared with bulk YBCO, all thin films in our samples are subject to an additional (biaxial) tensile strain due to the SrTiO$_3$ substrate \cite{StrainMultilayer}. This strain may undergo subtle variations between sample growth with and without the LCMO layer due to changes in the growth environment and relaxation within the LCMO layer. These considerations may help to rationalize why we could not find diffraction peaks associated with an electronic supermodulation (whether period-3 or period-4) on 30-nm YBCO thin films ($T_c\sim 60$ K) grown directly on the SrTiO$_3$ substrate. Incidentally, we note that no period-3 charge order has yet been reported on YBCO thin films, including Ref. \cite{CO_YBCO_Ghiringhelli} in which charge orders were found in NBCO thin films but not in YBCO. A systematic study of YBCO thin films with varying substrates and post-annealing conditions may provide a good means to elucidate the effects of strain on the charge ordering phenomena in YBCO, as well as whether other effects brought about by the LCMO layer are also essential to the manifestation of period-4 electronic supermodulation.

The aforementioned three factors are implicated with another phenomenon for which YBCO/LCMO heterostructures have been known--the long-range proximity effect between superconductivity and ferromagnetism as the two major competing orders in the system. As in other superconductor-ferromagnet heterostructures \cite{SF_Review}, superconductivity in YBCO and ferromagnetism in LCMO tend to suppress each other at the interface and extend their influence into the opposing layer. But because of the short superconducting coherence length along the c-axis in YBCO ($<$ 3 $\AA$), the high spin polarization (close to 100 \%) and large exchange splitting ($\sim$ 3 eV) in LCMO, the mutual suppressions of both orders in this system are expected to be limited to within a close vicinity of the interface ($\sim$ 1-nm away). Instead, several experiments have uncovered persistent effects that exist over a remarkably long range of several tens of nanometres \cite{SefriouiPRB, HoldenPRB, PenaPRB, StrainMultilayer, e-ph_longrange}. Understanding such an unusual effect requires consideration of factors beyond (singlet) superconductivity and ferromagnetism. While long-range effects caused by charge transfer \cite{ChargeXfer_longrange, ChargeXfer_shortrange}, strain \cite{StrainMultilayer}, disorder, defects \cite{IntergrownDefect} and triplet superconductivity \cite{TripletSC_PRB, TripletSC_NP} are under debate, a possible relevance of competing orders has been discussed very little, in part because existing candidates for the competing order, such as antiferromagnetism \cite{Chakhalian_NP} induced at the interface, are not sufficiently long ranged. Our observation of a new 3D long-range order inherent in YBCO/LCMO heterostructures that competes with superconductivity provides a new angle for addressing this problem: Superconductivity far from the interface can be directly suppressed by the electronic supermodulation even if ferromagnetism is absent or limited to the interfacial region. To what degree this mechanism is relevant to the observed long-range proximity effect remains to be seen.
\\

\noindent{\bf\large\textbf{Methods} }\\
\noindent{\bf Samples.}
Each of the studied YBCO thin films has 25 u.c. and a thickness of $\sim$ 30-nm, and was grown by pulsed laser deposition on top of a 25 u.c. ($\sim$ 10-nm) LCMO layer, which was initially deposited on the (001)-oriented SrTiO$_3$ (STO) substrate. A KrF ($\lambda$=248-nm) excimer laser, with 10 Hz repetition rate and 250 mJ power, was used to evaporate the target. The STO substrate was treated with HF-NH$_4$F buffer solution to produce a uniform TiO$_2$ termination. Direct deposition of YBCO/LCMO on top of the TiO$_2$-terminated STO makes a MnO$_2$-terminated interface (La$_{0.7}$Ca$_{0.3}$-MnO$_2$-BaO-CuO$_2$) between YBCO and LCMO (Sample \#2), whereas insertion of a buffer layer of 1.5 u.c. SrRuO$_3$ (SRO) between the STO substrate and LCMO layer produces a La$_{0.7}$Ca$_{0.3}$O-terminated interface (MnO$_2$-La$_{0.7}$Ca$_{0.3}$-CuO$_2$-BaO) (Sample \#1) \cite{Tra, Yu}. The growth temperature was set at 700 $^\circ$C and 715 $^\circ$C with an oxygen pressure of 80 mTorr and 200 mTorr for the deposition of LCMO and YBCO layers, respectively. All oxide layers were grown layer-by-layer as monitored \textit{in-situ} by reflection high-energy electron diffraction (Supplementary Fig. 1). An annealing treatment at 600 $^\circ$C with an oxygen pressure of 700 Torr was performed \textit{in situ} on each grown sample for 40 minutes followed by a slow cool down to room temperature to ensure full oxygenation. A capping layer of $\sim$ 10-nm LaAlO$_3$ was used to protect a portion of the sample surface for \textit{ex-situ} REXS measurements.
\\

\noindent{\bf Characterizations.}
Four-probe resistivity and SQUID magnetization measurements were performed on the sample soon after the growth. The superconducting transition temperature ($T_c$) of Samples \#1 \& \#2 are 55 $\pm 10$ K and 72 $\pm 7$ K, respectively. During the magnetization measurement, a 100 Oe magnetic field was applied perpendicular to the sample surface, and the field-cooled results are presented. The magnetic Curie temperature ($T_m$) was found to be the same for all samples studied, $\sim$ 175 K, irrespective of slight variations in the growth and annealing conditions. This value is similar to that reported earlier on YBCO/LCMO in Ref. \cite{Chakhalian_NP}.
\\

\noindent{\bf REXS.}
REXS experiments were mainly carried out in the scattering endstation at BL 4.0.2 of the Advanced Light Source (ALS). Main results have been reproduced in experiments performed in two additional scattering endstations located at BL 8.0.1 of the ALS and the REIXS beamline of the Canadian Light Source. Scattered intensity was measured using an in-vacuum CCD detector. CCD images were captured at a combination of detector and sample incidence angles, with each experimental geometry corresponding to a specified $(H, 0, L)$ reflection near the center of the CCD screen. Within each raw CCD image, the horizontal axis is parallel to the primary scattering plane, cutting through the diffraction peak in the $(H, 0, L)$ plane with a trajectory depending on the photon energy, and the sample and detector angles. The vertical axis of each image is perpendicular to the scattering plane and essentially parallel to the $[0, K, 0]$ direction. Fig. \ref{Fig. 1}c shows one such image projected onto the $(H, K, 0)$ plane. Intensity line cuts as a function of $H$ at fixed $K$, $L$ and photon energy (Figs. \ref{Fig. 1}d-f, \ref{Fig. 3}a \& c), as a function of $L$ at fixed $H$, $K$ and photon energy (Fig. \ref{Fig. 2}a-b) and as a function of photon energy at fixed $H$, $K$ and $L$ (Fig. \ref{Fig. 4}a), are extracted from a sequence of CCD images taken at different sample and detector angles with the center of each image corresponding to the designated $(H, 0, L)$. We note that when the diffraction condition is met for the supermodulation peak, its intensity is only in or around the central area of the CCD. Additionally reciprocal space maps constructed from the full CCD pixel array were used to confirm the 3D nature of the supermodulation (Supplementary Fig. 4 and Supplementary Note 1). In-plane lattice constant $a=3.855$ $\AA$ is assumed by averaging the a- and b-axis lattice constants reported on bulk and thin film samples at comparable doping levels \cite{CO_YBCO_Ghiringhelli}. The c-axis lattice constant $c=11.64$ $\AA$ is an averaged value of those determined from the $(0, 0, 1)$ (lattice) Bragg peak for samples at $\sim$50 K. Various factors contributed to the estimated experimental uncertainties in the supermodulation wavevector, including variations in the measured lattice constant among different experiments and as a function of temperature, imperfect sample-to-beam alignment, crystalline mosaicity, curvature of the thin films, etc.
\\

$^{*}$Corresponding authors: ruihua.he@bc.edu\\

\vspace{10mm}

\noindent {\bf\large Acknowledgement}\\
We thank M. Hashimoto for useful discussion. The work at Boston College was supported by a BC startup fund (J.F.H., R.-H.H.), US NSF CAREER Award DMR-1454926 (R.-H.H., in part), NSF Graduate Research Fellowship GRFP-5100141 (T.R.M.) and NSF MRI grant DMR-1337576 (M.J.G.). The work in NCTU is supported by Ministry of Science and Technology, R.O.C. (MOST 103-2119-M-009-003-MY3), Center for Interdisciplinary Science of National Chiao Tung University, Ministry of Education, Taiwan (MOE-ATU 101W961). Research was mainly performed at the ALS, which is supported by the Director, Office of Science, Office of Basic Energy Sciences, of US DOE under Contract No. DE-AC02-05CH11231. Research was partially performed at the Canadian Light Source (proposal \#19-5803), which is funded by the CFI, NSERC, NRC, CIHR, the Government of Saskatchewan, WD Canada, and the University of Saskatchewan.
\\

\noindent {\bf\large Author Contributions}\\
J.F.H., P.S. and T.M. led the REXS experiments with R.-H.H. and Q.H.. V.T.T., J.Y.L., and Y.H.C. grew and characterized the samples. P.S., Y.-D.C., W.L.Y. and E.A. maintained the REXS endstations. J.F.H., P.S., and R.-H.H. analyzed the data with assistance from J.K.. J.F.H., P.S. and R.-H.H. wrote the paper with suggestions and comments by E.A. and T.M.. R.-H.H., E.A., Y.-H.C., J.Y.L., and M.J.G. were responsible for project direction, planning and infrastructure.
\\

\noindent {\bf\large Additional information}\\
\noindent{\bf Competing financial interests:} The authors declare no competing financial interests.

\newpage

\end{document}